\documentclass[conference]{IEEEtran}
\IEEEoverridecommandlockouts
\usepackage{cite}
\usepackage{graphicx}
\usepackage{textcomp}
\usepackage[draft]{hyperref}

\usepackage{amsmath,amssymb,amsfonts}
\usepackage{algpseudocode}
\usepackage{algorithm}
\usepackage[table,xcdraw]{xcolor}

\newcommand\tab[1][1cm]{\hspace*{#1}}
\def\BibTeX{{\rm B\kern-.05em{\sc i\kern-.025em b}\kern-.08em
    T\kern-.1667em\lower.7ex\hbox{E}\kern-.125emX}}

\begin{document}

\title{Dynamic Role-Based Access Control for Decentralized Applications\\
}

%---------TB uncommented-----------------------------------

% \author{\IEEEauthorblockN{Arnab Chatterjee, Yash Pitroda}
% \IEEEauthorblockA{\textit{Robert Bosch Engineering and Business Solutions Private Limited} \\
% Bengaluru, India \\
% \{arnab.chatterjee, yash.pitroda\}@bosch.com}
% }

\author{
    \IEEEauthorblockN{Arnab Chatterjee\IEEEauthorrefmark{1}, Yash Pitroda\IEEEauthorrefmark{1}, Manojkumar Parmar\IEEEauthorrefmark{1} \IEEEauthorrefmark{2}}
    \IEEEauthorblockA{\IEEEauthorrefmark{1} \textit{Robert Bosch Engineering and Business Solutions Pvt. Ltd.} \\ Bengaluru, India
    \\\{arnab.chatterjee,yash.pitroda,manojkumar.parmar\}@bosch.com}
    \IEEEauthorblockA{\IEEEauthorrefmark{2} \textit{HEC Paris}\\ Jouy-en-Josas Cedex, France
    \\manojkumar.parmar@hec.edu}
}

%--------------------------------------------
\maketitle

\begin{abstract}
Access control management is an integral part of maintaining the security of an application. Although there has been significant work in the field of cloud access control mechanisms, however, with the advent of Distributed Ledger Technology (DLT), on-chain access control management frameworks hardly exist. Existing access control management mechanisms are tightly coupled with the business logic, resulting in governance issues, non-coherent with existing Identity Management Solutions, low security, and compromised usability.\\
We propose a novel framework to implement dynamic role-based access control for decentralized applications (dApps). The framework allows for managing access control on a dApp, which is completely decoupled from the business application and integrates seamlessly with any dApps. The smart contract architecture allows for the independent management of business logic and execution of access control policies. It also facilitates secure, low cost, and a high degree of flexibility of access control management. The proposed framework promotes decentralized governance of access control policies and efficient smart contract upgrades. This paper also provides performance comparison with adjacent access control frameworks for DLT. Any Turing complete smart contract programming language is an excellent fit to implement the framework. We expect this framework to benefit enterprise and non-enterprise dApps and provide greater access control flexibility and effective integration with traditional and state of the art identity management solutions.
\end{abstract}


\section{Introduction}
\label{introduction}
For the past few years, Distributed Ledger Technology (DLT) and decentralized applications (dApps) have revolutionized how businesses are imagined. The proliferation of enterprise systems bring sophistication into modern day to day transactions. Transactions today are bound across multiple systems. Conventional information security practices are challenging to adhere to while building and maintaining such complex systems. DLT promises to simplify some of these sophistications. \\
\tab DLT acts as a shared, immutable ledger between these systems, offering a single source of truth for exchanged data. However, concerns around information security, especially data confidentiality laws and regulations with based DLT systems have become one of the disinclinations towards enterprise DLT adoption \cite{23_Chat1909_Production} \cite{02_BlockchainDisillusionmentGartner}.  
Two of the core defining characteristics of information security triad (Confidentiality, Integrity, and Availability) are confidentiality and integrity \cite{01_ferraiolo_kuhn_chandramouli_2007}, access control is critical to preserve these characteristics.  
Over the last few decades, there have been several advances in the area of access control mechanisms. \\
\tab Access control defines and constrains what a user can do in a system. In other words, it authorizes the user for certain activities that s/he wishes to perform. Although, access control can take numerous forms \cite{01_ferraiolo_kuhn_chandramouli_2007} in addition to what users can do, like when and how the resources might be used, manual interventions like voting before granting access to specific resources. \\
\tab Computation, whether a particular user has certain rights to access resources, usually happens in a siloed and insecure computational context in current cloud/on-premise solutions. With the advent of DLT, secure multiparty computation is now possible backed by consensus among participants on the outcome of the computation. Since data and code both reside on an immutable shared ledger, secure, tamper-proof computation is burgeoning. Access control of shared data on a DLT is essential due to compliance with information security needs, as described above. Two of the widely accepted and used Blockchain platforms, namely Hyperledger Fabric\cite{03_DBLP:journals/corr/abs-1801-10228} and Ethereum\cite{04_wood2014ethereum}, supports access control mechanisms in smart contracts. However, they fall short in the following areas:

\begin{enumerate}
\item \textbf{Tight coupling with the source code of the contract } \\
Roles often need to be tightly bound with the functionalities \cite{11_accesscontrol-openzeppelin} \cite{09_RBAC} \cite{10_ABAC}.

\item \textbf{Lacks dynamism} \\
Upgrading roles, granting and revoking permissions for a role is often cumbersome, leading to source code change and smart contract redeployment, as role-based or attribute-based access control is often performed inside the target function. For example, modifiers in solidity \cite{17_dannen2017introducing} are applied at the target source code site, i.e. along with the function definition.

\item \textbf{Governance Chaos} \\
Governance mechanisms need to be re-executed every time on the deployment of the updated contract. For example, a change in either of roles, permissions, or policy, the entire business logic needs to be reaudited, consented with stakeholders and redeployed. These unnecessary changes result in chaos in governing the change management process \cite{18_chohan2017decentralized}. 

\item \textbf{Cost} \\
Redeployment of contracts leads to gas fees (in some cases), and expensive security audits\cite{16_destefanis2018smart}.

\item \textbf{Security} \\
Security vulnerabilities may arise in executing the new contract on old data with new permissions \cite{19_luu2016making}\cite{21_parizi2018smart}.

\item \textbf{Integration with Identity Management Suite} \\
Integration and interoperability with cloud Identity Management Solutions (IdMs) becomes complicated as each cloud provider has non-standard IdM suites with different role definitions and attributes \cite{20_yasin2016online}.

\item \textbf{Performance Issues} \\
Existing solutions are expensive in nature \cite{11_accesscontrol-openzeppelin} as they involve nested calls back and forth with policy execution contracts.

\item \textbf{Compromised Usability} \\
Development usability of the contract is reduced as the developer often has to worry about defining the correct roles in the function definition\cite{21_parizi2018smart}.

\end{enumerate}

This paper consists of the following sections. Section \ref*{problem_statement} describes the problem statement formally. Section \ref*{our_contribution} introduces our contribution. Section \ref*{dynamic_role_based_access_control} talks about the solution called \textit{Dynamic Role-Based Access Control}, and introduces the mathematical primitives of the problem domain and describes the policy execution for access control. Section \ref*{solution_architecture} discusses the solution architecture and contextualizes its solution using Unified Modelling Language(UML) notation. Section \ref*{discussion} discusses, interprets and describes the significance of the findings and explores the underlying meaning of this solution. Section \ref*{future_work} closes the paper with the possible future directions to enhance the solution. Finally, Section \ref*{summary} summarises the problem and the solution. 

\section{Problem Statement}
\label{problem_statement}
One of the significant challenges of implementing access control mechanisms is that the landscape of the business is continually changing\cite{05_bauer2009real}\cite{06_fundamentals_of_information_systems_security_access_control_systems}. 
Different use-cases require specific accesses control mechanisms based on roles of users and business workflows. Implementation access control in a cross-organisational shared ledger requires dynamism to accommodate these rapidly changing needs. Since Smart Contract is the executable logic that resides on the immutable ledger, access control requires to be implemented here.\\
\tab The smart contract logic is often meant to focus on operations and core business logic on the shared data. This shared data is stored as a global state along with the operations that can be performed on it. In reality, the semantics of today’s smart contract languages allow access control to applied at a much lower level in the business logic itself. Change in access control rules requires touching business logic code. For example, modifiers in Solidity\cite{07_dannen_2017} are applied while defining a function or using attribute-based access control in Hyperledger Fabric chaincode allows performing checks about the target function level \cite{08_gaur_desrosiers_ramakrish_2018}.\\
\tab This intertwined identity and access control code with the business logic allows for less flexibility to change the access control rules or the business logic independently. This kind of tightly coupled \cite{14_martin2002agile} code leads to violation of the single responsibility principle \cite{14_martin2002agile}. As a consequence, several problems like increased cost of redeployments and audits, approval, governance and voting chaos, performance, usability, and other issues (as described above) arise. 

\section{Our Contribution}
\label{our_contribution}
This solution framework described in this paper (henceforth referred to as \textit{solution}) aims to tackle the problems described above. We tackle to problem of high coupling \cite{14_martin2002agile} by using a permissions manager contract to check the user details, the role s/he possesses, the function s/he is trying to access, and based on a set of dynamically configured rules, the call is delegated to the target contract at runtime. The business logic is decoupled from the access control code. Smart contracts can be dynamically added, and the permissions manager contract can be configured at runtime with the whereabouts, the functions definitions, and the roles that can access the respective functions\\
\tab The solution also allows provisioning for dynamically adding users, user-role mappings, and function-role mappings. This sort of loosely coupled \cite{14_martin2002agile} design allows for dynamic upgrade \cite{15_lin2017survey} and configuration of smart contracts and roles without the additional cost, governance problems, performance, and usability issues.

\section{Dynamic Role-Based Access Control}
\label{dynamic_role_based_access_control}
Dynamic Role-Based Access Control is inspired by PERM Modelling Language\cite{12_luo2019pml}. It provides a generic solution framework to define permissions in a smart contract. The mathematical primitives of the entity sets like roles, users, policies and resources, are first defined in an agnostic manner such that it fits any smart contract language irrespective of the blockchain platform. Next, the management of these entity sets are described. Finally, how these entity sets are used to execute a policy at runtime is elaborated.

\subsection{Primitives}
The following section defines the primitives used to define role based access control through the rest of the paper.

\begin{enumerate}

\item A \textbf{User} is an entity and the set $U=\{u_1,u_2,…,u_n \}$ is a finite set of $n$ users 

\item A \textbf{Role} is an entity and the set $R=\{r_1,r_2,…,r_p \}$ is a finite set of $p$ roles 

\item A \textbf{Function} is an entity and the set $F=\{f_1,f_2,…,f_q \}$ is a finite set of $q$ functions 

\item The relation between $F$ and $R$ is defined by $F_R: R\xrightarrow{}F $, a many-to-many mapping consisting of the functions that can be accessed by different roles and is called the \textbf{function-role mapping}

\item The relation between $U$ and $R$ is defined by $U_R : U \xrightarrow{} R$ a many-to-many mapping consisting of the roles that users possess and is called the \textbf{user-role mapping}
 
\item A request $Req$ consists of a tuple $Req = (u_i, f_j)$ where $u_i \in U, \forall i=\{1,2,...,n\} $ and $f_j \in F, \forall j=\{1,2,...,q\} $

\end{enumerate}

\iffalse
\subsection{Management}
This section deals with the management of entity sets $U$,$R$, and $F$ and mappings $U_R$ and $U_R$. It includes how the different entities are created/onboarded; their lifecycle is managed viz. detail updated, and finally, how these entities are deleted/disposed of from the system. The management functions isolate the entities from the policy logic and allow for easy change management.  In the following sub-sections we will define the entities and the Application Programming Interfaces (APIs)  in an Object-Oriented Programming style using a class diagram (refer figure \ref*{fig:class_diagram}).
\fi

\begin{figure}
  \includegraphics[width=\linewidth]{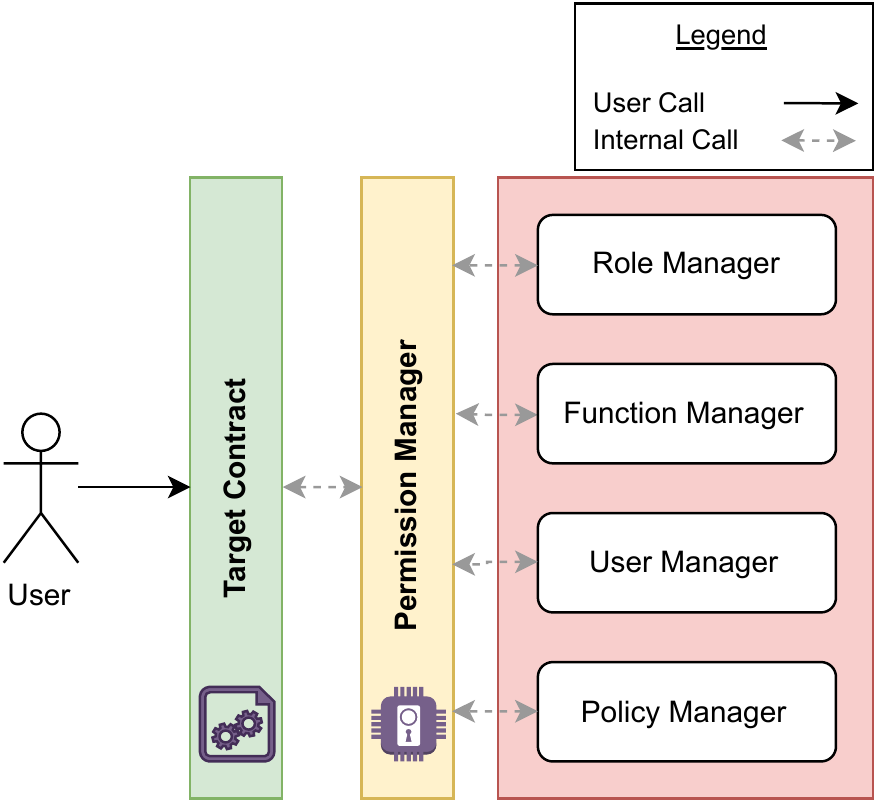}
  \caption{Block Diagram of the Solution Framework}
  \label{fig:block_diagram}
\end{figure}

\begin{figure*}
  \includegraphics[width=\linewidth]{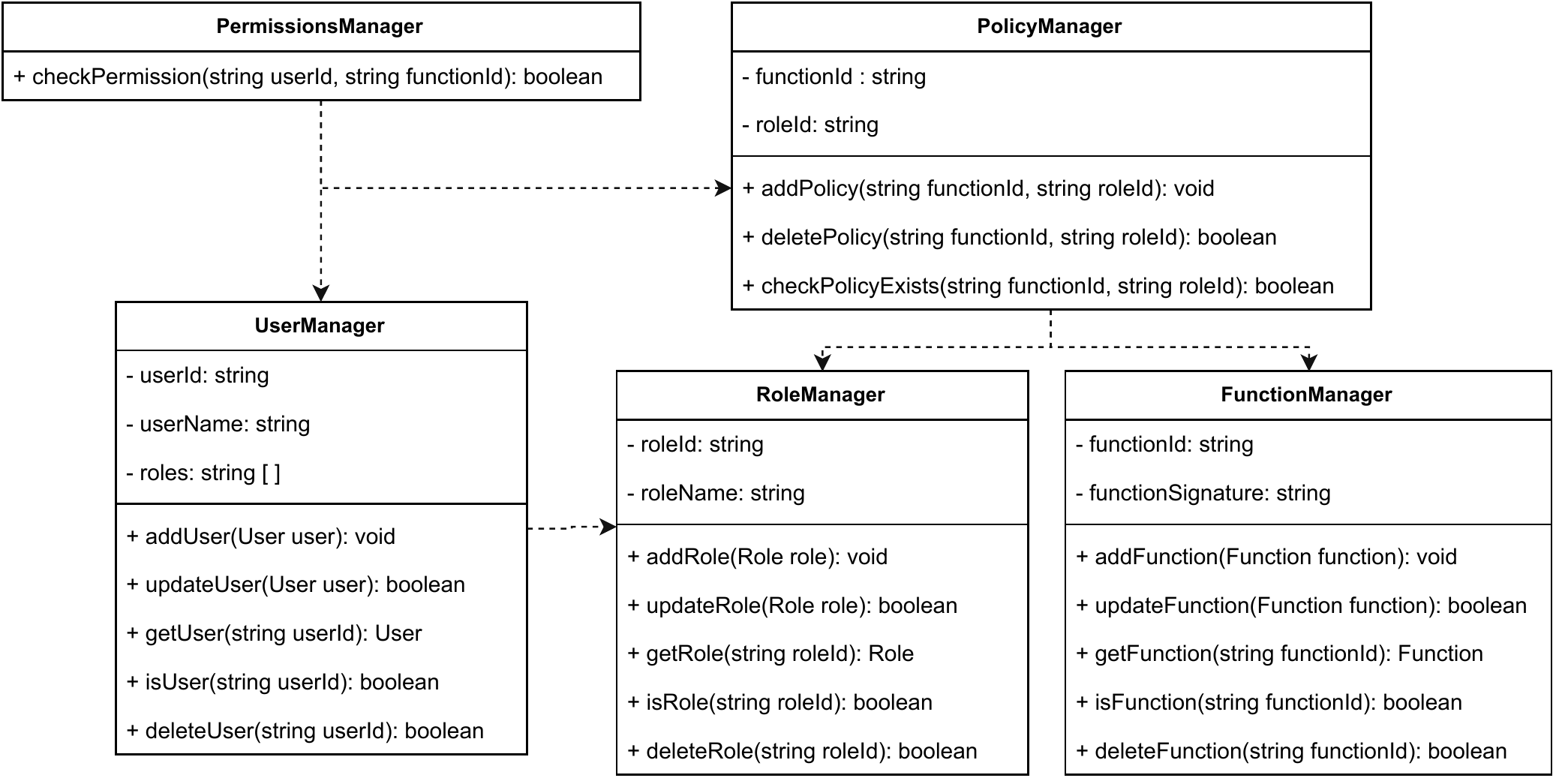}
  \caption{UML Class Diagram of the Solution Framework}
  \label{fig:class_diagram}
\end{figure*}

\subsection{Policy Execution}

The policy is executed whenever a user tries to access a function in the contract. The request defined by $Req$ arrives at the system; the target contract accepts the request and delegates it to the Permission Manager which breaks down the request tuple attributes into the $u_i$ and $f_j$.\\

If $u_i \in U$, we find the range/co-domain of $u_i$ from $U_R$. Let the range be $r_i \subseteq R $.\\

If  $f_j \in F$, we find the range/co-domain of $f_j$ from $F_R$. Let the range be $r_j \subseteq R$.\\

To execute the policy, we simply check $r_i \cap r_j \neq \phi $. 

\vspace{4pt}
\section{Solution Architecture}
\label{solution_architecture}
We implement the mathematical concepts demonstrated previously using UML. 

The block diagram of the solution is represented in figure \ref*{fig:block_diagram}. It shows a layered diagram of the different manager components forming the architecture of the system. 
The description of each of these manager components are described below.

\begin{enumerate}
    \item \textbf{Role Manager}: Manages roles in the system, i.e. $R$. 
    \item \textbf{Function Manager}: Manages functions in the system, i.e. $F$
    \item \textbf{User Manager}: Manages users and roles possessed by the users i.e., $U$ and $U_R$
    \item \textbf{Policy Manager}: Manages roles that are eligible to access a function, i.e., $F_R$ 
    \item \textbf{Permissions Manager}: Orchestrates the flow among the Role, Function, User and Policy Manager, validate the flow and output a boolean response. 
\end{enumerate}

The manager components cater to the following requirements:
\begin{enumerate}

    \item Adding, updating, fetching, checking existence and disposing of different entities from the system, viz. roles, functions, users and policies.
	
	\item Maintains consistency of data. For example, any role that is linked to any user or function cannot be disposed or altered accidentally.
	
	\item 	Allows for capturing comprehensive information about the entities Role, Function, User, and Policy. For example the Policy Manager can decide if an user is required to have $m$ out of $p$ roles to grant access to a function. 
	
	\item Decoupled flow control aiding in smoother management and governance. For example, in case of a change in role management logic, only the role component can be replaced with the data of roles intact. This type of decoupling can be achieved using \textit{Eternal Storage Design Pattern} \cite{13_eternal_storage_solidity-patterns} in Solidity programming language.
    
    \item The decoupled approach also leads to optimized deployment and storage costs
	
	\item Independent and Fine grained control over each management component. For example: A group of super administrators, say “Group A” can manage Roles and Users, while “Group B” can manage Policies, and finally “Group C” can vet and approve the changes. Such complex governance is cumbersome and challenging to achieve using conventional access control techniques.
\end{enumerate}

We define the entities and the Application Programming Interfaces (APIs) for the described manager components in Object-Oriented Programming (OOP) paradigm using a class diagram, as shown in figure \ref*{fig:class_diagram}.\\

\begin{figure*}
  \includegraphics[width=\linewidth]{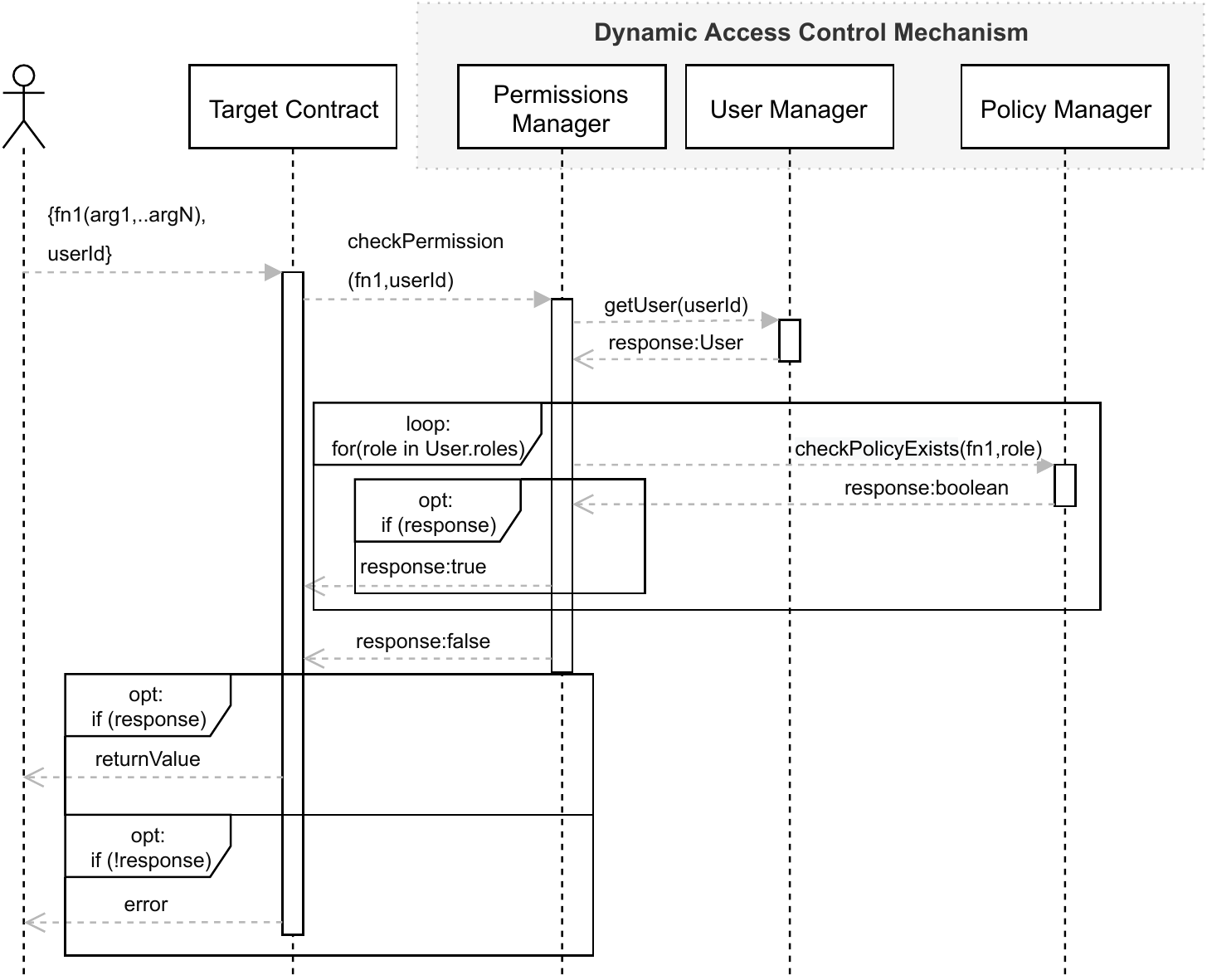}
  \caption{UML Sequence Diagram of the Solution Framework}
  \label{fig:sequence_diagram}
\end{figure*}

\tab The sequence of events that follow the policy execution described above is depicted in the sequence diagram in figure \ref*{fig:sequence_diagram}. 

\begin{enumerate}
    \item The user invokes a function $f_j$ with certain arguments. The user identity $u_i$ also flows along and lands in the target smart contract. 
    
    \item The target contract delegates these arguments to the Permissions Manager
    
    \item The Permissions Manager gets the details of the user $u_i$, more specifically the roles the user possess (say $r_i$)
    
    \item The Permissions Manager checks is any policy mapping exists with the Policy manager by passing the function $f_j$ and the roles $r_i$
    
    \item The Policy Manager iterates over the set of roles assigned to $f_j$ (say  $r_j$) and returns \textit{true} if a match exists between the sets $r_i$ and  $r_j$ or \textit{false} otherwise
    
    \item The decision is propagated back to the target contract, where it resumes execution in case of a Boolean \textit{true} response or throws an Authorization error otherwise
\end{enumerate}

For the sake of brevity, the following checks from the request $Req$  are omitted from the sequence diagram: 

\begin{enumerate}
    \item 	Existence of user $u_i \in U$ 
	\item Existence of function $f_j \in F$
	\item Existence of role $r_p \in R$

\end{enumerate}

% Description 

\section{discussion}
\label{discussion}
The solution approaches the well-known problem of role-based access control from the perspective of a decentralized context. It takes into consideration that multiple parties involved in the ledger are unknown, and each resource or function has a set of well-defined access-control rules called policies against which parties have access permissions to perform specific actions. It offers a dynamic perspective, considering the need for flexibility and control over authorization.\cite{23_Chat1909_Production}. It also looks at the need for having a mechanism to update the data and the actions on the data (code) in a flexible manner and avoiding substantial code changes every time. This mechanism is particularly useful when the deployed code, as access control rules are subject to change, mainly for the two reasons. First is the dynamic landscape of business, and second is the lack of mature formal verification methods of smart contracts\cite{22_bhargavan2016formal}.\\
\tab Moreover, since ledger writes can only be updated and not reverted or deleted. The logical separation of code or actions and the data is necessary. The loose coupling demonstrated of each of the entities shown the Solution Architecture discussed in the paper allows for such flexibility and precise control over access control. \\
\tab Additionally, the approach described aids in implementing a decentralized change management  process\cite{18_chohan2017decentralized}. For example, a voting mechanism can be implemented to allow specific changes in policies to take effect, with zero change in disjoint contracts or logic, also leaving less room for any chaos in governance. The approach also leads to better usability of the core business logic contracts allowing them to change independently, without the developer having to worry about the roles and permissions allowed for the given smart contract function. \\
\tab The current approach supports implementation in Generation 2 blockchain solutions like Ethereum and Hyperledger Fabric. We have implemented the solution in Solidity programming language (version 0.5.11) for Ethereum Virtual Machine. Below are some quantitative comparisons of our implementation vis-à-vis OpenZeppelin’s implementation for Access Control\cite{11_accesscontrol-openzeppelin}. Below are two important observations:
\begin{enumerate}
    \item 	Unit Deployment costs for our solution are higher than OpenZeppelin’s solution. 
However, when we compare it to the number of upgrades (say $N_{UG}$) in the overall lifecycle of a contract, due to changes in business logic, access control, users, and policies in an enterprise context, our solution is deployed once and hence economical. However, the solution from OpenZeppelin requires re-deployment for every change in role or policies leading to higher long term costs.  Moreover, OpenZeppelin requires a new contract to be deployed every time a new role is added into the system, thereby making the contracts, even more, costlier for multiple roles than our solution. 
Table \ref*{table_costs} summarises the approximate costs of deploying the same contract for one role and $N_R$ roles. It also compares the cost of upgrading the contract $N_{UG}$ times. \\
\tab We can, therefore, see the costs for our solution are constant if we extrapolate it for $N_R$ roles and  $N_{UG}$ upgrades, whereas it increases linearly for Open Zeppelin’s solution. \\
\\

\item Our solution fares better for long term transaction cost efficiency. One time transaction Costs (Gas Used) for our solution are approximately 40\% higher than OpenZeppelin’s solution. However, the cost for OpenZeppelin's solution increases linearly when more roles are added, and complex computation needs to be executed when checking role access to a function. The lack of dynamism in OpenZeppelin's solution as discussed in Section \ref{introduction} is chiefly responsible for this linear increase in transaction cost. In contrast to this, the transaction costs for our solution viz. costs for checking access remains constant even when scaled for multiple role checks. 

\begin{table}[]
\caption{Deployment Cost Comparison for Our Solution vs. OpenZeppelin's Solution}
\label{table_costs}
\begin{tabular}{|l|l|l|}
\hline
\rowcolor[HTML]{C0C0C0} 
\textbf{} & \textbf{Our Solution} & \textbf{\begin{tabular}[c]{@{}l@{}}OpenZeppelin’s\\ Solution\end{tabular}} \\ \hline
\begin{tabular}[c]{@{}l@{}}One Time Deployment Cost\\ for One Role and Zero \\ Upgrades (Gas Used)\end{tabular} & $\sim$9536190 & $\sim$359268 \\ \hline
\begin{tabular}[c]{@{}l@{}}Deployment costs for \\ $N_R$ roles witg $N_{UG}$ \\ upgrades (Gas Used)\end{tabular} & $\sim$9536190 & \begin{tabular}[c]{@{}l@{}}$\sim$359268 x $N_R$ x $N_{UG}$\end{tabular} \\ \hline
\end{tabular}
\end{table}

\end{enumerate}

\section{Future Work}
\label{future_work}
Some of the means how the described solution is extensible in the future are as follows:

\begin{enumerate}
    \item The policy manager is extensible to evaluate sophisticated policies. For example, the policy manager can be improvised to consider the roles and permissions of the organization while evaluating the policies at runtime.
    \item The change mechanism can be extended to include voting to enable decentralized decision making for change management\cite{18_chohan2017decentralized}.
    \item  The current approach requires the manual management of functions and users in the function manager and user manager respectively. This approach can be automated. Automatically synchronization is possible in this case. For example, using an identity management provider, the user details are synchronized to the user manager. Likewise, functions in the smart contract are automatically synchronized using a source code parser to extract and smart contract functions and synchronize with the function manager.
    \item More quantitative information about the performance and security of the described approach can be provided. This would help decision-makers evaluate the performance and security of the solution before deploying it.
    \item Ethereum smart contract implementations can be optimized for Gas Costs. The current implementation does not focus on gas optimization best practices.
\end{enumerate}

\iffalse
\begin{algorithm}
	\caption{Policy Manager} 
	\begin{algorithmic}[1]
	    \Function{GetFunctionRoles}{$F$}
	       \State roles = []
	       \For {\text{\textbf{each} role \textbf{r} in $r$}}
	            \If{\text{r} $\Longleftrightarrow$ \text{F exist in $R$}}
	                \State \text{roles.add(r)}
	            \EndIf
	       \EndFor
	       \State \Return roles
    	\EndFunction
	\end{algorithmic} 
\end{algorithm}

\begin{algorithm}
	\caption{User Manager} 
	\begin{algorithmic}[1]
	    \Function{GetUserRoles}{$U$}
	       \State userRoles = [ ]
	       \For {\text{\textbf{each} role \textbf{r} in $r$}}
	            \If{\text{U} $\Longleftrightarrow$ \text{r exist in $UR$}}
	                \State \text{userRoles.add(r)}
	            \EndIf
	       \EndFor
	       \State \Return userRoles
    	\EndFunction
	\end{algorithmic} 
\end{algorithm}

\begin{algorithm}
	\caption{Authorizer} 
	\begin{algorithmic}[1]
	    \Function{checkAuthorization}{$F,U$}
	        \State funcRoles = \Call{GetFunctionRoles}{$F$}
    	    \State userRoles = \Call{GetUserRoles}{$U$}
    	    \For {\text{\textbf{each} r in funcRoles}}
    	        \If {\text{r in userRoles}}
    	            \State \Return $True$
    	        \EndIf
    	    \EndFor
    	    \State \Return $False$
    	\EndFunction
	\end{algorithmic} 
\end{algorithm}

\fi

\section{Summary}
\label{summary}
In this paper we have demonstrated an approach to solve the problem of decentralized role-based access control. The demonstrated mechanism is loosely coupled \cite{14_martin2002agile} with the business logic allowing both the access control policies and the business logic to change independently. This segregation of concerns allows for the following benefits: easy and independent auditing and governance of smart contract and its deployment, reduced costs of change management, easier integration with cloud identity management. On a bigger scale the mentioned approach is also easier manage. We have also demonstrated a few observations with our implementation compared to standard industry libraries for access control. The presented solution framework is suited for use in both enterprise and public dApp context. We expect this framework to enable flexible access control management in dApps and address specific challenges for enterprise DLT adoption.

%\vspace{12pt}

\bibliographystyle{IEEEtran}
\bibliography{bibliography}

\begin{thebibliography}{10}
\providecommand{\url}[1]{#1}
\csname url@samestyle\endcsname
\providecommand{\newblock}{\relax}
\providecommand{\bibinfo}[2]{#2}
\providecommand{\BIBentrySTDinterwordspacing}{\spaceskip=0pt\relax}
\providecommand{\BIBentryALTinterwordstretchfactor}{4}
\providecommand{\BIBentryALTinterwordspacing}{\spaceskip=\fontdimen2\font plus
\BIBentryALTinterwordstretchfactor\fontdimen3\font minus
  \fontdimen4\font\relax}
\providecommand{\BIBforeignlanguage}[2]{{%
\expandafter\ifx\csname l@#1\endcsname\relax
\typeout{** WARNING: IEEEtran.bst: No hyphenation pattern has been}%
\typeout{** loaded for the language `#1'. Using the pattern for}%
\typeout{** the default language instead.}%
\else
\language=\csname l@#1\endcsname
\fi
#2}}
\providecommand{\BIBdecl}{\relax}
\BIBdecl

\bibitem{23_Chat1909_Production}
A.~Chatterjee, M.~S. Parmar, and Y.~Pitroda, ``Production challenges of
  distributed ledger technology {(DLT)} based enterprise applications,'' in
  \emph{2019 International Symposium on Systems Engineering (ISSE) (IEEE ISSE
  2019)}, Edinburgh, United Kingdom (Great Britain), Sep. 2019.

\bibitem{02_BlockchainDisillusionmentGartner}
\BIBentryALTinterwordspacing
S.~Haig, ``Blockchain enters "trough of disillusionment" according to
  gartner.'' [Online]. Available:
  \url{https://news.bitcoin.com/blockchain-enters-trough-disillusionment-gartner}
\BIBentrySTDinterwordspacing

\bibitem{01_ferraiolo_kuhn_chandramouli_2007}
D.~Ferraiolo, D.~R. Kuhn, and R.~Chandramouli, \emph{Role based access
  control}.\hskip 1em plus 0.5em minus 0.4em\relax Artech House, 2007.

\bibitem{03_DBLP:journals/corr/abs-1801-10228}
\BIBentryALTinterwordspacing
E.~Androulaki, A.~Barger, V.~Bortnikov, C.~Cachin, K.~Christidis, A.~D. Caro,
  D.~Enyeart, C.~Ferris, G.~Laventman, Y.~Manevich, S.~Muralidharan, C.~Murthy,
  B.~Nguyen, M.~Sethi, G.~Singh, K.~Smith, A.~Sorniotti, C.~Stathakopoulou,
  M.~Vukolic, S.~W. Cocco, and J.~Yellick, ``Hyperledger fabric: {A}
  distributed operating system for permissioned blockchains,'' \emph{CoRR},
  vol. abs/1801.10228, 2018. [Online]. Available:
  \url{http://arxiv.org/abs/1801.10228}
\BIBentrySTDinterwordspacing

\bibitem{04_wood2014ethereum}
G.~Wood \emph{et~al.}, ``Ethereum: A secure decentralised generalised
  transaction ledger,'' \emph{Ethereum project yellow paper}, vol. 151, no.
  2014, pp. 1--32, 2014.

\bibitem{11_accesscontrol-openzeppelin}
\BIBentryALTinterwordspacing
 [Online]. Available:
  \url{https://docs.openzeppelin.com/contracts/2.x/access-control}
\BIBentrySTDinterwordspacing

\bibitem{09_RBAC}
J.~P. {Cruz}, Y.~{Kaji}, and N.~{Yanai}, ``Rbac-sc: Role-based access control
  using smart contract,'' \emph{IEEE Access}, vol.~6, pp. 12\,240--12\,251,
  2018.

\bibitem{10_ABAC}
E.~{Yuan} and J.~{Tong}, ``Attributed based access control (abac) for web
  services,'' in \emph{IEEE International Conference on Web Services
  (ICWS'05)}, July 2005, p. 569.

\bibitem{17_dannen2017introducing}
C.~Dannen, \emph{Introducing Ethereum and Solidity}.\hskip 1em plus 0.5em minus
  0.4em\relax Springer, 2017.

\bibitem{18_chohan2017decentralized}
U.~W. Chohan, ``The decentralized autonomous organization and governance
  issues,'' \emph{Available at SSRN 3082055}, 2017.

\bibitem{16_destefanis2018smart}
G.~Destefanis, M.~Marchesi, M.~Ortu, R.~Tonelli, A.~Bracciali, and R.~Hierons,
  ``Smart contracts vulnerabilities: a call for blockchain software
  engineering?'' in \emph{2018 International Workshop on Blockchain Oriented
  Software Engineering (IWBOSE)}.\hskip 1em plus 0.5em minus 0.4em\relax IEEE,
  2018, pp. 19--25.

\bibitem{19_luu2016making}
L.~Luu, D.-H. Chu, H.~Olickel, P.~Saxena, and A.~Hobor, ``Making smart
  contracts smarter,'' in \emph{Proceedings of the 2016 ACM SIGSAC conference
  on computer and communications security}.\hskip 1em plus 0.5em minus
  0.4em\relax ACM, 2016, pp. 254--269.

\bibitem{21_parizi2018smart}
R.~M. Parizi, A.~Dehghantanha \emph{et~al.}, ``Smart contract programming
  languages on blockchains: An empirical evaluation of usability and
  security,'' in \emph{International Conference on Blockchain}.\hskip 1em plus
  0.5em minus 0.4em\relax Springer, 2018, pp. 75--91.

\bibitem{20_yasin2016online}
A.~Yasin and L.~Liu, ``An online identity and smart contract management
  system,'' in \emph{2016 IEEE 40th Annual Computer Software and Applications
  Conference (COMPSAC)}, vol.~2.\hskip 1em plus 0.5em minus 0.4em\relax IEEE,
  2016, pp. 192--198.

\bibitem{05_bauer2009real}
L.~Bauer, L.~F. Cranor, R.~W. Reeder, M.~K. Reiter, and K.~Vaniea, ``Real life
  challenges in access-control management,'' in \emph{Proceedings of the SIGCHI
  Conference on Human Factors in Computing Systems}.\hskip 1em plus 0.5em minus
  0.4em\relax ACM, 2009, pp. 899--908.

\bibitem{06_fundamentals_of_information_systems_security_access_control_systems}
``Fundamentals of information systems security/access control systems,'' in
  \emph{Fundamentals of Information Systems Security/Access Control Systems -
  Wikibooks, open books for an open world}.

\bibitem{07_dannen_2017}
C.~Dannen, \emph{Introducing ethereum and solidity: foundations of
  cryptocurrency and blockchain programming for beginners}.\hskip 1em plus
  0.5em minus 0.4em\relax Apress, 2017.

\bibitem{08_gaur_desrosiers_ramakrish_2018}
N.~Gaur, L.~Desrosiers, V.~Ramakrishna, N.~Petr, S.~A. Baset, and A.~ODowd,
  \emph{Hands-on blockchain with Hyperledger: building decentralized
  applications with Hyperledger Fabric and composer}.\hskip 1em plus 0.5em
  minus 0.4em\relax Packt Publishing, 2018.

\bibitem{14_martin2002agile}
R.~C. Martin, \emph{Agile software development: principles, patterns, and
  practices}.\hskip 1em plus 0.5em minus 0.4em\relax Prentice Hall, 2002.

\bibitem{15_lin2017survey}
I.-C. Lin and T.-C. Liao, ``A survey of blockchain security issues and
  challenges.'' \emph{IJ Network Security}, vol.~19, no.~5, pp. 653--659, 2017.

\bibitem{12_luo2019pml}
Y.~Luo, Q.~Shen, and Z.~Wu, ``Pml: An interpreter-based access control policy
  language for web services,'' 2019.

\bibitem{13_eternal_storage_solidity-patterns}
\BIBentryALTinterwordspacing
``Eternal storage.'' [Online]. Available:
  \url{https://fravoll.github.io/solidity-patterns/eternal_storage.html}
\BIBentrySTDinterwordspacing

\bibitem{22_bhargavan2016formal}
K.~Bhargavan, A.~Delignat-Lavaud, C.~Fournet, A.~Gollamudi, G.~Gonthier,
  N.~Kobeissi, N.~Kulatova, A.~Rastogi, T.~Sibut-Pinote, N.~Swamy
  \emph{et~al.}, ``Formal verification of smart contracts: Short paper,'' in
  \emph{Proceedings of the 2016 ACM Workshop on Programming Languages and
  Analysis for Security}.\hskip 1em plus 0.5em minus 0.4em\relax ACM, 2016, pp.
  91--96.

\end{thebibliography}

\end{document}